\documentclass[12pt]{iopart}

\newcommand{\eeq}{\end{equation}}
\newcommand{\beq}{\begin{equation}}
\newcommand{\ba}{\begin{array}}
\newcommand{\ea}{\end{array}}
\newcommand{\bea}{\begin{eqnarray}}
\newcommand{\eea}{\end{eqnarray}}
\newcommand{\vev}[1]{\langle #1\rangle}

\newcommand{\heth}{${^3}$He}
\newcommand{\hefo}{${^4}$He}
\newcommand{\lisi}{${^6}$Li}
\newcommand{\lise}{${^7}$Li}
\newcommand{\bese}{${^7}$Be}

\begin{document}

\title[BBN and fundamental ``constants'']{Big bang nucleosynthesis as a probe of fundamental ``constants''}

\author{Thomas Dent, Steffen Stern and Christof Wetterich}

\address{Theoretische Physik, Universit{\"a}t Heidelberg, Philosophenweg 16, 69120 Heidelberg Germany}
\ead{T.Dent@ThPhys.Uni-Heidelberg.de}

\begin{abstract}
Big Bang nucleosynthesis (BBN) is the earliest sensitive probe of the values of many fundamental particle physics parameters. We have found the leading linear dependences of primordial abundances on all relevant parameters of the standard BBN code, including binding energies and nuclear reaction rates. This enables us to set limits on possible variations of fundamental parameters. We find that ${^7}$Li is expected to be significantly more sensitive than other species to many fundamental parameters, a result which also holds for variations of coupling strengths in grand unified (GUT) models. Our work also indicates which areas of nuclear theory need further development if the values of ``constants'' are to be more accurately probed. 
\end{abstract}

%\section{Motivation and method}
\noindent The constancy over space and time of the coupling strengths and particle masses in the Standard Model of particle physics is an assumption that should be tested \cite{Uzanreview}. Variations may arise in a relativistically covariant theory due to the coupling of Standard Model particles to a scalar field whose cosmological value depends on time \cite{Bekenstein82}. The possibility of time-varying couplings has been discussed %\cite{Wetterich88_2} 
since the first suggestions of dynamical dark energy \cite{Wetterich88RatraPeebles} due to a scalar field evolving over recent cosmological epochs.

Further motivation comes from possible signals of nonzero variation at redshifts $0.5$--$4$, arising from absorption spectra which probe the fine structure constant $\alpha$ and the proton-electron mass ratio $\mu\equiv m_p/m_e$.
%, and the proton gyromagnetic ratio $g_p$. 
The observational situation is contradictory, with both nonzero \cite{Murphy03,Reinhold06} and null \cite{Levshakov} %,Tzanavaris06} 
results, and debate about methods and statistical and systematic errors \cite{Murphycrit}. 
Bounds at different redshifts, and in different environments, probe possible variations without relying on particular models. We consider variations at the epoch of BBN ($z \simeq 10^{10}$), which is currently the earliest time at which theories of nuclear and particle physics can be compared to astrophysical observation.
%\cite{PDGFields,Steigman05,Sarkar95}.

The study of particle physics via BBN faces two theoretical challenges. First, there are many variable parameters, to be compared with the small number of reliably observable abundances. Second, the uncertainty in how QCD parameters, in particular quark masses, affect nuclear forces, and thus nuclear binding energies and cross-sections. A derivation of nuclear forces from first principles, apart from the long-range attraction attributable to pion exchange, is lacking.
%\footnote{Recent efforts in lattice QCD \cite{Aokinuclear} have been encouraging.
Instead, various types of effective theories have been used which fit measured nuclear properties.

We present a systematic approach to these challenges. Firstly, each variation in particle physics or ``fundamental'' parameters is considered independently. Thus at linear order we can allow for {\em any}\/ theoretical scenario in which such variations are subject to some unified relation. Second, we identify which nuclear properties and reactions have significant influence on the variation of the primordial abundances. This is done by varying every relevant nuclear binding energy and cross-section in a modified reaction integration code 
%\cite{Wagoner69,Wagoner72,Kawano88,Kawano92} 
and noting the dependence of output abundances. 
%(Changes to the code are discussed in more detail in Appendix A.) 
This response to ``nuclear'' parameters is then to be related to the variation of fundamental parameters via nuclear theory. The connection between the different levels of physical understanding arises by simple matrix multiplication of ``response matrices''.

The dependence of primordial abundances $Y_a$ with $a=($D, \heth, \hefo, \lisi, \lise) on the variation of a set of nuclear physics parameters $X_i$ is given by a response matrix $C$ with matrix elements \cite{MSW}
\beq \label{Cdef}
 c_{ai} = \frac{\partial \ln Y_a}{\partial \ln X_i}.
\eeq
extracted by considering small variations of the $X_i$ within the BBN code. This includes variation of the reaction rates which have a physical dependence on $X_i$. 
% All variations in parameters are taken to be small.
%, such that all necessary information can indeed be extracted from the response matrix. 
Then, the nuclear physics parameters $X_i$ are related to a set of Standard Model parameters $G_k$ via a second response matrix $F$ with entries
\beq \label{Fdef}
 f_{ik} = \frac{\partial \ln X_i}{\partial \ln G_k}.
\eeq
This step requires, at present, theoretical assumptions, and contains substantial uncertainties in the dependence of nuclear binding energies on quark masses, which we discuss following Equations \ref{BDmhat} and \ref{dBdmpi}.
%, which we discuss in Section~\ref{fundamentals}. 
The variation of abundances with respect to fundamental parameters $G_k$ is then given by a matrix $R$, obtained simply via $R=CF$, with elements $r_{ak}$:
\beq
 \frac{\Delta Y_a}{Y_a} = r_{ak} \frac{\Delta G_k}{G_k}.
\eeq
%The matrix $R$ is then obtained by simple matrix multiplication: .
We use mass-energy units where the QCD invariant scale $\Lambda_c$ 
%(sometimes written $\Lambda_{\rm QCD}$) 
is kept constant. This is convenient for nuclear reactions where mass scales are determined mainly by strong interactions. Variations of dimensionful parameters are measured relative to $\Lambda_c$, for example considering the electron mass $m_e$ we implement a variation of $m_e/\Lambda_c$. 
%\paragraph{Nuclear parameters}

We vary with respect to the following thirteen ``nuclear'' parameters $X_i$: %in the BBN code:
\begin{itemize}
\item Gravitational constant $G_N$
\item Neutron lifetime $\tau_n$
\item Fine structure constant $\alpha$
\item Electron mass $m_e$
\item Average nucleon mass $m_N\equiv(m_n+m_p)/2$
\item Neutron-proton mass difference $Q_N\equiv m_n-m_p$
%\item Reaction rate, for each nuclear reaction affecting species with $A\leq 7$
\item Binding energies of D, T, ${^3}$He, ${^4}$He, \lisi, \lise, ${^7}$Be.
\end{itemize}
%Clearly to vary each binding energy independently is unphysical, however our purpose is to determine the leading (linear) variation of all abundances, with respect to the code's input parameters, once and for all. Then given any specific theoretical model, we can construct a linear combination of variations of $X_i$
% nuclear parameters 
%to account for any variation of a fundamental parameter. 
Our results for these parameters are shown in Table \ref{dlnYdlnX}, where the rows constitute the transposed nuclear response matrix $C^T$; detailed discussion can be found in \cite{us07}.
%We also quote the dependence of the abundances on $\eta$ in the last row. Values are quoted to 2 d.\,p.\ or to 2 sig.\ fig.\ when the magnitude exceeds 1. Below we give a few comments concerning specific parameters $X_i$.
\begin{table}
\centering
\caption{Response matrix $C$, dependence of abundances on nuclear parameters.} \label{dlnYdlnX}
\vspace*{2mm}
\begin{tabular}{|c|c|c|c|c|c|}
\hline
$\partial\ln Y_a/\partial\ln X_i$  & D & \heth & \hefo & \lisi & \lise \\
\hline \hline
%\endhead
$G_N$         &  0.94 &  0.33 &  0.36 &  1.4  & -0.72 \\ \hline
$\alpha$      &  2.3  &  0.79 &  0.00 &  4.6  & -8.1  \\ \hline
$\tau_n$      &  0.41 &  0.15 &  0.73 &  1.4  &  0.43 \\ \hline
$m_e$         & -0.16 & -0.02 & -0.71 & -1.1  & -0.82 \\ \hline
$Q_N$         &  0.83 &  0.31 &  1.55 &  2.9  &  1.00 \\ \hline
$m_N$         &  3.5  &  0.11 & -0.07 &  2.0  &-12    \\ \hline
$B_{\rm D}$   & -2.8  & -2.1  &  0.68 & -6.8  &  8.8  \\ \hline
$B_{\rm T}$   & -0.22 & -1.4  &  0    & -0.20 & -2.5  \\ \hline
$B_{\rm 3He}$ & -2.1  &  3.0  &  0    & -3.1  & -9.5  \\ \hline
$B_{\rm 4He}$ & -0.01 & -0.57 &  0    &-59    &-57    \\ \hline
$B_{\rm 6Li}$ &  0    &  0    &  0    & 69    &  0    \\ \hline
$B_{\rm 7Li}$ &  0    &  0    &  0    &  0    & -6.9  \\ \hline
$B_{\rm 7Be}$ &  0    &  0    &  0    &  0    & 81    \\ \hline
%\hline
%$\eta$        & -1.6  & -0.57 &  0.04 & -1.5  &  2.1  \\ \hline
%$G_N$         &  0.94 &  0.33 &  0.36 &  1.40 & -0.72 \\ \hline
%$\alpha$      &  2.33 &  0.79 &  0    &  4.60 & -8.10 \\ \hline
%$\tau_n$      &  0.41 &  0.15 &  0.73 &  1.37 &  0.43 \\ \hline
%$m_e$         & -0.16 & -0.02 & -0.71 & -1.11 & -0.82 \\ \hline
%$Q_N$         &  0.83 &  0.31 &  1.55 &  2.88 &  1.00 \\ \hline
%$m_N$         &  3.50 &  0.11 & -0.07 &  1.99 &-11.93 \\ \hline
%$B_{\rm D}$   & -2.81 & -2.11 &  0.68 & -6.83 &  8.83 \\ \hline
%(see below)   & -1.51 & -2.41 &  0.66 & -5.55 &  7.51 \\ \hline
%$B_{\rm T}$   & -0.22 & -1.42 &  0    & -0.20 & -2.54 \\ \hline
%$B_{\rm 3He}$ & -2.15 &  3.00 &  0    & -3.07 & -9.54 \\ \hline
%$B_{\rm 4He}$ & -0.01 & -0.57 &  0    &-58.6  &-56.75 \\ \hline
%$B_{\rm 6Li}$ &  0    &  0    &  0    & 68.85 &  0    \\ \hline
%$B_{\rm 7Li}$ &  0    &  0    &  0    &  0    & -6.90 \\ \hline
%$B_{\rm 7Be}$ &  0    &  0    &  0    &  0    & 81.26 \\ \hline
%$\eta$        & -1.59 & -0.57 &  0.04 & -1.50 &  2.08 \\ \hline
\end{tabular}
\vspace*{-3mm}
\end{table}
The largest sensitivity of abundances to nuclear parameters involves the variation of \hefo, \lisi\ and \bese\ binding energies. For instance the rate for \heth$(\alpha,\gamma)$\bese\ with $Q$-value 1.59\,MeV is very sensitive to changes in these (numerically large) binding energies. $Q$-values affect abundances both by influencing the reverse reaction rate as
% are simply related to the forward rate via statistical factors, thus
%, due to time reversal invariance (see for example \cite{NACRE}): the relevant dependence is  
\beq
\frac{\vev{\sigma v}_{34\rightarrow 12}}{\vev{\sigma v}_{12\rightarrow 34}} \propto e^{-Q/T}
\eeq
and via the kinematic dependence of cross-sections. For dipole radiative capture reactions we have
%(assuming a dominant electric dipole) 
\beq \label{radiative}
\sigma(E) \propto E_\gamma^3 \sim (Q+E)^3
\eeq
whereas for $2\rightarrow 2$ inelastic scattering or transfer reactions the dependence is
\beq \label{2to2}
\sigma(E) \propto \beta \sim (Q+E)^{1/2}
\eeq
where $\beta$ is the outgoing channel velocity. For resonances, we scale their contributions to thermal reaction rates, which vary as $e^{-E_r/T}$, by the appropriate power of $(Q+E_r)$. 

The dependence of reaction matrix elements on binding energies and on $Q$ is in general not clear due to the lack of systematic effective theory. For the $npd\gamma$ reaction we use the nuclear effective theory result of \cite{Chen99} where dependence on $B_{\rm D}$ is explicit.
In order to diagnose which reaction rates are important in the variation of final abundances, we varied each thermal averaged cross-section $\vev{\sigma v}$ by a temperature-independent factor, preserving the relation between forward and reverse rates. Apart from $npd\gamma$ and $n\leftrightarrow p$, only seven reaction rates strongly influence final abundances (discounting \lisi). In every case the dependences $\partial\ln Y_a/\partial\ln \vev{\sigma v}_i$ are order unity or smaller \cite{us07}.
%, and for many ``important'' reactions only a few abundances are significantly affected. 
Also, the \hefo\ fraction does not depend on any reaction cross-section apart from $n\leftrightarrow p$.

%\section{Fundamental parameters and GUT models}
Next we connect these nuclear parameters to fundamental parameters $G_k$ at a higher energy scale. We consider the following six fundamental parameters: 
%\begin{itemize}
%\item 
the gravitational constant $G_N$; 
%\item 
the fine structure constant $\alpha$; 
%\item 
the electron mass $m_e$;
%\item 
the light quark mass difference $\delta_q \equiv m_d-m_u$; 
%\item 
the averaged light quark mass $\hat{m}\equiv (m_d+m_u)/2 \propto m_\pi^2$; and
%\item 
the Higgs v.e.v.\ $\vev{\phi}$. These are to be varied independently.
%\end{itemize}
The strange quark mass $m_s$ is omitted from our list because the present theoretical uncertainties in its influence on nuclear parameters are too high. 

Linear variations in the nuclear parameters $X_i$ are then encoded in the matrix $F$ defined in (\ref{Fdef}): our estimates of $F$ \cite{us07} are shown in Table \ref{dlnXdlnG}.
\begin{table}
\centering
\caption{Response matrix $F$, dependence of nuclear parameters $X_i$ on fundamental parameters $G_k$.}\label{dlnXdlnG}
\vspace*{2mm}
\begin{tabular}{|c||c|c|c|c|c|c|}
\hline 
$\partial \ln X_i/\partial\ln G_k$ & $G_{\rm N}$ & $\alpha$ & $\vev{\phi}$ & $m_e$ & $\delta_q$ & $\hat{m}$ \\
\hline \hline
$G_{\rm N}$ & 1 & 0 & 0 & 0 & 0 & 0 \\
\hline 
$\alpha$ & 0 & 1 & 0 & 0 & 0 & 0 \\
\hline
$\tau_n$ & 0 & 3.86 & 4 & 1.52 & -10.4 & 0 \\
\hline
$m_e$ & 0 & 0 & 0 & 1 & 0 & 0 \\
\hline
$Q_N$ & 0 & -0.59 & 0 & 0 & 1.59 & 0 \\
\hline
$m_N$ & 0 & 0 & 0 & 0 & 0 & 0.048 \\
\hline
$B_{\rm D}$ & 0 & -0.0081 & 0 & 0 & 0 & $-4$ \\
\hline
$B_{\rm T}$ & 0 & -0.0047 & 0 & 0 & 0 & $-2.1 f_{\rm T}$ \\
\hline
$B_{3\rm He}$ & 0 & -0.093 & 0 & 0 & 0 & $-2.3 f_{3\rm He}$ \\
\hline
$B_{4\rm He}$ & 0 & -0.0304 & 0 & 0 & 0 & $-0.94 f_{4\rm He}$ \\
\hline
$B_{6\rm Li}$ & 0 & -0.0541 & 0 & 0 & 0 & $-1.4 f_{6\rm Li}$ \\
\hline
$B_{7\rm Li}$ & 0 & -0.0459 & 0 & 0 & 0 & $-1.4 f_{7\rm Li}$ \\
\hline
$B_{7\rm Be}$ & 0 & -0.0885 & 0 & 0 & 0 & $-1.4 f_{7\rm Be}$ \\
\hline
\end{tabular}
\vspace*{-3mm}
\end{table}
We estimate the leading dependence of nuclear binding energies on the pion mass, due to the pion-mediated contribution to the nucleon-nucleon potential, 
% which are dominant within the expectation values of the interaction potentials used \cite{Pieper01, Pudliner97}, 
by scaling up the dependence of the deuteron binding energy. 
%\cite{EpelbaumMeissner02,BeaneSavage02}. 
For the deuteron, 
\beq \label{BDmhat}
 \Delta \ln B_{\rm D} \equiv r \Delta \ln m_\pi = \frac{r}{2} \Delta \ln \hat{m}
\eeq
with $r\approx -8 \pm 2$ \cite{YooScherrer}.
%$-10\leq r \leq -6$.
%\footnote{Our definition of $r$ differs by a sign from \cite{YooScherrer}.}
Then for a nucleus with mass number $A_i$ we estimate
\beq \label{dBdmpi}
 \frac{\partial B_i}{\partial m_\pi} = f_i (A_i-1) \frac{B_{\rm D}}{m_\pi} r \simeq
 -0.13 f_i (A_i-1) 
\eeq
where the constants $f_i$ are expected to be of order unity, but may vary in magnitude by factors of a few. This is our main source of theoretical uncertainty, in the case where $\hat{m}$ varies significantly. However, we expect our main finding of a much larger dependence of \lise\ than of D and \hefo\ to be robust, barring fine-tuned cancellations. 
Table~\ref{dlnYdlnG} then shows the dependences of abundances on fundamental parameters, encoded in the matrix $R$. 
\begin{table}
\centering
\caption{Response matrix $R$, dependence of abundances $Y_i$ on fundamental parameters $G_k$.}\label{dlnYdlnG}
\vspace*{2mm}
\begin{tabular}{|c||c|c|c|c|c|}
\hline 
$\partial \ln Y_a/\partial\ln G_k$ &  D   & \heth & \hefo & \lisi  & \lise \\ \hline \hline
$G_{\rm N}$                        & 0.94 & 0.33  & 0.36  &  1.4   & -0.72 \\ \hline 
$\alpha$                           & 3.6  & 0.95  & 1.9   &  6.6   &-11    \\ \hline
$\vev{\phi}$                       & 1.6  & 0.60  & 2.9   &  5.5   &  1.7  \\ \hline
$m_e$                              & 0.46 & 0.21  & 0.40  &  0.97  & -0.17 \\ \hline
$\delta_q$                         &-2.9  &-1.1   &-5.1   & -9.7   & -2.9  \\ \hline
$\hat{m}$                          &17    & 5.0   &-2.7   & -6     &-61    \\ \hline
\hline
$\eta$                             &-1.6  &-0.57  & 0.04  & -1.5   &  2.1  \\ \hline
\end{tabular}
\vspace*{-3mm}
\end{table}
For the $\hat{m}$-dependences
%, which involve the nuclear binding energies with their uncertain values of $f_i$, 
we have given the values which arise when setting all $f_i$ to unity. 

We may now set bounds on the variation of each fundamental parameter, considered in isolation. Out of three observational determinations of primordial abundances (D, \hefo\ and \lise) the observed \lise\ abundance is a factor two to three smaller than standard BBN theory predicts. % (SBBN).
% and systematic uncertainties related to stellar evolution exist \cite{Korn06}. 
Thus we use D and \hefo\ for the purpose of constraining allowed variations.
For deuterium we take $2 \sigma$ limits from \cite{OMeara06};
% demand that the abundance should not deviate by more than  from the observed value
for \hefo\ we consider instead the ``conservative allowable range'' of \cite{OliveSkillman04}. The resulting constraints are given in Table~\ref{AllowedVariations}.
\begin{table}
\centering
\caption{Allowed individual variations 
%($2\sigma$ or ``conservative allowable range'') 
of fundamental couplings.}\label{AllowedVariations}
\vspace*{2mm}
\begin{tabular}{|clcc|}
\hline
$-19 \% $&$\le\ \Delta \ln G_N        $&$\le$&$ +10 \%  $\\ \hline
$-3.6\% $&$\le\ \Delta \ln \alpha     $&$\le$&$ +1.9\%  $\\ \hline
$-2.3\% $&$\le\ \Delta \ln \vev{\phi} $&$\le$&$ +1.2\%  $\\ \hline
$-17 \% $&$\le\ \Delta \ln m_e        $&$\le$&$ +9.0\%  $\\ \hline
$-0.7\% $&$\le\ \Delta \ln \delta_q   $&$\le$&$ +1.3\%  $\\ \hline
$-1.3\% $&$\le\ \Delta \ln \hat{m}    $&$\le$&$ +1.7\%  $\\
\hline
\end{tabular}
\vspace*{-3mm}
\end{table}

We also consider unified scenarios where the variations of fundamental couplings satisfy relations that reduce the number of free parameters. If there 
%of a parameter $G_k$ 
is a single underlying degree of freedom which varies, written as a dimensionless scalar $\varphi$, then variations of $G_k$ can then be written as a vector:
%\beq
$ \Delta \ln G_k = d_k \Delta \varphi, $
%\eeq
where 
% which gives rise to the variation and 
$d_k$ are a set of numbers characterising a particular unified model. We then obtain
\beq
 \Delta \ln Y_a = (CF)_{ak} d_k \Delta \varphi.
\eeq
One may eliminate $\Delta \varphi$ in favour of the variation of another parameter, such as $\alpha$.
We considered a grand unified theory with unified gauge coupling $\alpha_X$, broken at the scale $M_X$ to the Standard Model group. The observable couplings of QCD and electromagnetism are related to $\alpha_X$ via renormalization group flow and electro\-weak sym\-metry-breaking. We take for simplicity the Planck mass fixed relative to the unification scale, $\Delta (M_{\rm P}/M_X)=0$, and the Yukawa couplings to be constant, thus electron and quark masses are proportional to $\vev{\phi}$. The variation of $\vev{\phi}$ is parameterised as
\beq
 \frac{\vev{\phi}}{M_X} = \mbox{const.}\left(\frac{\Lambda_c}{M_X}\right)^{\gamma}.
\eeq
We found two cases of interest. In the first, the Higgs v.e.v.\ 
%and elementary fermion masses are all 
is proportional to the unification scale, thus 
%$\Delta(M_X/M_{\rm P},\vev{\phi}/M_{\rm P},m_{e,q}/M_{\rm P})=0$, or equivalently 
$\gamma=0$. Then we find \cite{us07}
\beq \label{dGGUT2}
 \Delta \ln(G_{\rm N},\alpha,\vev{\phi},m_e,\delta_q,\hat{m}) \simeq (64.5,1,-32,-32,-32,-32) \Delta \ln \alpha. 
\eeq
We then obtain variations of abundances
\beq
 \Delta \ln (Y_{\rm D}, Y_{3\rm He}, Y_p,
%{4\rm He}
Y_{6\rm Li}, Y_{7\rm Li}) \simeq 
 (-450, -130, 170, 380, 1960) \Delta \ln \alpha.
\eeq
Note the strong variation of \lise\ compared to other abundances. Here, a fractional variation of $\alpha$ by about $-2.5\times 10^{-4}$
%$0.025 \%$  %({\it i.e.}\ a fractional variation of ) 
would bring abundances within 2$\sigma$ observational bounds.
%, while remaining in the linear regime. Conversely, in the ``GUT3'' model 
Secondly, we consider a case when the Higgs v.e.v.\ and fermion masses vary {\em more}\/ rapidly (with respect to $M_X$) than the QCD scale: thus $\gamma > 1$. We take $\gamma = 1.5$ and find variations of fundamental couplings
\beq \label{dGGUT3}
 \Delta \ln(G_{\rm N},\alpha,\vev{\phi},m_e,\delta_q,\hat{m}) \simeq (87,1,21.5,21.5,21.5,21.5) \Delta \ln \alpha. 
\eeq
The variations of abundances are then
\beq
 \Delta \ln (Y_{\rm D}, Y_{3\rm He}, Y_p, 
%{4\rm He} 
Y_{6\rm Li}, Y_{7\rm Li}) \simeq 
 (430,130,-65,-60,-1420) \Delta \ln \alpha.
\eeq
In this scenario a fractional variation of $\alpha$ by about $+4\times 10^{-4}$
%$0.04 \%$, {\it i.e.}\ $\Delta \ln \alpha = , 
brings theory and observation into agreement within 1$\sigma$ bounds. Thus the ``lithium problem'' may be considerably ameliorated by allowing variable ``constants''. 

In both scenarios, as \lise\ decreases, \hefo\ also decreases, but the D abundance increases. Then we would expect the true primordial \hefo\ fraction to be at the lower side of its currently allowed range, while D is near its upper observational limit.

%\section{Conclusion}

In summary, primordial nucleosynthesis is a unique window on the very early universe, setting the earliest or most stringent bounds on many effects in cosmology and particle physics. The possibility that particle masses and couplings could have a space-time dependence can thus be tested, and we present a systematic treatment of potential signals and bounds at BBN. In general there are more variable parameters than observables, but if unification of gauge couplings is assumed, definite limits emerge, and preferred models may be found where the predicted \lise\ abundance is strongly suppressed, while the helium fraction decreases and the deuterium abundance increases, within current observational limits.

\end{document}